# Multi-Octave Metamaterial Reflective Half-Wave Plate for Millimetre and Sub-Millimetre wave Applications

Giampaolo Pisano,[1,*] Bruno Maffei,[2] Peter A.R. Ade,[1] Paolo de Bernardis,[3] Peter de Maagt,[4] Brian Ellison,[5] Manju Henry,[5] Ming Wah Ng,[1] Brian Schortt,[6] Carole Tucker[1]

[1]School of Physics and Astronomy, Cardiff University, CF24 3AA Cardiff, UK
[2]IAS, Université Paris-Sud, Orsay, 91405, France
[3]Dipartimento di Fisica, Universita' di Roma La Sapienza, 00185 Roma, Italy
[4]Electromagnetics & Space Environments Division, European Space Agency, NL 2200 AG Noordwijk, The Netherlands
[5]Rutherford Appleton Laboratory, Harwell Oxford, OX11 0QX Didcot, UK
[6]Future Missions Office, Science Directorate, European Space Agency, NL 2200 AG Noordwijk, The Netherlands
*Corresponding author: giampaolo.pisano@astro.cf.ac.uk

The quasi-optical modulation of linear polarization at millimeter and sub-millimeter wavelengths can be achieved by using rotating half wave plates (HWPs) in front of polarization sensitive detectors. Large operational bandwidths are required when the same device is meant to work simultaneously across different frequency bands. Previous realizations of half wave plates, ranging from birefringent multi-plate to mesh-based devices, have achieved bandwidths of the order of 100%. Here we present the design and the experimental characterization of a reflective HWP able to work across bandwidths of the order of 150%. The working principle of the novel device is completely different from any previous realization and it is based on the different phase-shift experienced by two orthogonal polarizations respectively reflecting off an electric conductor and off an artificial magnetic conductor.

## 1. Introduction

In the field of millimetre and sub-millimetre wave astronomy polarimetric rotating half-wave plates (HWPs) are often used to rotate of the angle of the incoming linear polarization. This has multiple advantages: the modulation of the polarized signals outside the 1/f noise of the instrument, the possibility to control the systematics of the optical components following the HWP and the ability to measure polarization along specific directions, a desirable requirement in scanning strategies.

This work is related to the technology development for the next generation of Cosmic Microwave Background (CMB) radiation satellites, funded by the European Space Agency. This project aims to develop large radii half-wave plates with the goal of achieving the optical performances required by a satellite similar to the one proposed by the COrE team [1].

On the one hand, many different technologies have already been employed to realise transmissive HWPs, which range from multi-plate birefringent designs [2] to metamaterial based solutions [3, 4, 5]. On the other hand, the CORE instrument required a large reflective HWP as the first optical element. Specifically, this instrument was meant have a large (1.2 m diameter) reflective HWP rotating half-wave plate based on a free-standing wire-grid (Fig.1).

The COrE instrument was intended to cover a large frequency range, from 60 to 600 GHz. This was achieved by a reflective HWP having narrow multiple periodic bands at the price of a lower overall efficiency. The challenges related to the development of such a large device triggered the development of a completely different solution which is based on the mesh filters technology [6, 7].

The novel design is based on a completely different working principle. The differential phase-shift between ordinary and extraordinary axes is neither achieved by mimicking a birefringent material, nor by using Pancharatnam designs, nor capacitive or inductive grids and nor by a displacing wire-grid.

The new concept relies on the difference in the reflection phases between a perfect electric conductor surface (PEC) and an artificial magnetic conductor (AMC) surface. The latter has been developed recently by using the mesh filters technology [8]. The bandwidth achieved by the new polypropylene embedded reflective HWP (ER-HWP) is superior to any other device ever realised in this field (Fig.2).

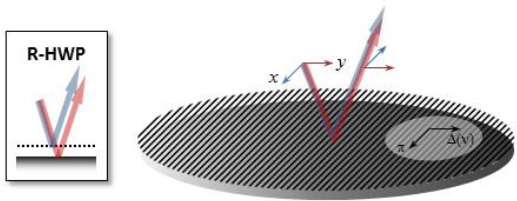

Fig. 1. Air-gap reflective HWP

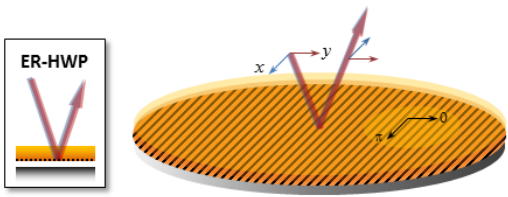

Fig. 2. Dielectrically embedded reflective HWP

## 2. Concept

In this section we will compare and contrast different solutions by using one of the most important performance parameters, the polarisation modulation efficiency. Assuming almost unitary reflection coefficients, we can define it as $\varepsilon = sin^2(\Delta\phi/2)$, where $\Delta\phi$ is the differential phase-shift introduced by the HWP between the two orthogonal axes. Ideal modulation efficiencies are achieved when the HWP induces a differential phase-shift $\Delta\phi = \pi$.

Transmission line (TL) modelling will be used to investigate normal incidence performance whereas more accurate finite element analysis [9] will be necessary in the off-axis cases and when metamaterial structures are adopted. The goal is to achieve high modulation efficiency across large bandwidths, low losses and by keeping the same performances when the device is used at oblique incidence (angular stability).

### A. Traditional air-gap reflective HWP (R-HWP)

The traditional air-gap reflective HWP constitutes of a free-standing wire-grid positioned parallel to a mirror at a specific distance, $d = \lambda_0/4$, where $\nu_0 = c/\lambda_0$ is the fundamental frequency [Fig. 3(a)]. The working principle is very simple. The polarisation component parallel to the grid is reflected whereas the orthogonal one goes through it, is reflected by the mirror and then recombines with the other polarisation. The extra path, equal to half a wavelength, adds a phase-shift equal to $\pi$ at the frequency $\nu_0$. We have chosen $\nu_0 = 30$ GHz for our models.

The same phase-shift is also achieved at odd multiples of the fundamental frequency, $\nu_n = (2n + 1)\nu_0$. However, at intermediate frequencies the phase-shift will deviate linearly from the ideal value and the modulation efficiency will vary sinusoidally, as shown in Fig. 4(a). This means that the plate is very efficient only within narrow and periodic frequency bands and on average only 50% of the linearly polarised radiation will be modulated.

The manufacture of large diameter wire-grids is very challenging. The mechanical structures holding the wires have to be very strong and massive. This problem led us investigate the following other options.

### B. Single-layer embedded R-HWP

It is possible to realise a simple embedded R-HWP based on the mesh-filters technology. The wire-grid can be replaced by the photolithographic version which is made by etching narrow copper strips on a polypropylene substrate. The substrate provides also the required quarter-wavelength $\lambda_0/4n$ which is, in this case, within the dielectric medium with refractive index $n$ [Fig. 3(b)].

The presence of the dielectric substrate creates a mismatch for the radiation propagating in free-space. The overall effect is to decrease the modulation efficiency down to ~40% by narrowing down the periodic bands [Fig. 4(b)]. In addition, at off-axis incidence angles the modulation efficiencies for the S- and P- polarisations, respectively orthogonal and parallel to the plane of incidence, will not be equal as a function of frequency.

### C. Two-layer embedded R-HWPs

It is possible to reduce the mismatch introduced by the dielectric substrate discussed in part B by adding an anti-reflection coating layer [Fig. 3(c)]. This layer will have a lower refractive index equal to $n_{ARC} = \sqrt{n}$ and thickness equal to $\lambda_0/4n_{ARC}$. As shown in Fig. 4(c), in this case the overall efficiency goes back to almost ~50%.

We could add more coating layers in the attempt to improve the efficiency but, even in the perfectly matched case, the efficiency would not exceed 50% because it would be like having the free-standing R-HWP completely embedded into a different medium. We are then going to investigate a completely different approach.

### D. Multi-layer embedded R-HWP

What we are trying to achieve is an ideal surface providing a differential phase-shift of $\pi$ along two orthogonal directions. For the polarisation parallel to the wires, the wire-grid behaves almost like a perfect conductor (PEC) surface by providing an almost frequency-independent phase-shift of $\pi$. What is needed is something producing a null phase-shift in the direction orthogonal to the wires. This could be achieved in principle if the device looked like an ideal perfect magnetic conductor (PMC) surface along that direction.

Even if PMC surfaces do not exist in nature, their behaviour is mimicked within certain bandwidths by the so called Artificial Magnetic Conductor (AMC) surfaces [10]. Recently, a very large bandwidth AMC based on the mesh-filters technology has been developed at millimetre wavelengths [8]. In this device the null phase-shift is achieved by using the internal reflection in a high-to-low refractive index interface. The radiation gets into the high index medium by means of a broadband antireflection coating. See reference [8] for a detailed analysis.

In the R-HWP design, a null phase-shift along the direction orthogonal to the wires can be then achieved if the radiation is gently brought into a high index medium and is then abruptly reflected off a high-to-low index interface. This can be achieved by a structure of the type sketched in Fig. 3(d). There are three layers with increasing refractive index, from $n_1 \cong 1.25$ to $n_3 \cong \sim 3$, followed by a lower index layer $n_4 \cong 1.5$ that could in principle be free-space.

Notice that the polarisation parallel to the wires will get into the high index medium and will be reflected by the wires as before experiencing a $\pi$ phase-shift (PEC). The orthogonal polarisation will be almost completely reflected on the same plane by the high-to-low index interface which will provide a null phase-shift (AMC) over a large bandwidth. Differently from the previous cases, both polarisation will now go through the same number of dielectric layers and will be reflected off at the same plane [Fig. 2].

The polarisation modulation efficiency greatly increases up to an average of 80% between successive maxima. Within the regions around the peaks the efficiency is very high, around 99%, and very flat [Fig. 4(d)]. This is due to the flatness of the differential phase-shift achieved within the same bands.

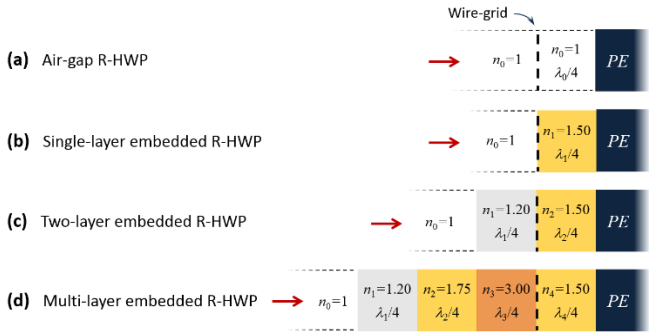

Fig. 3. Sketches of the different types of reflective HWPs.

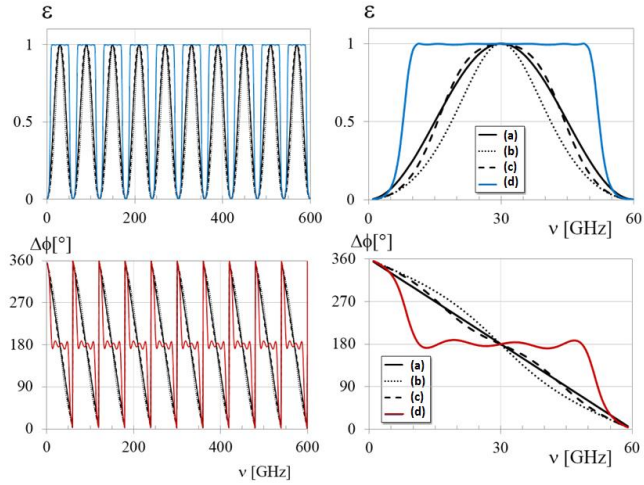

Fig. 4. Modulation efficiency and differential phase-shift for the different types of reflective HWPs (a), (b), (c) and (d) sketched in Fig. 3.

## 3. Design

In this section we are going to describe more in details the design of the embedded R-HWP based on the multi-layer AMC structure.

### A. Transmission line modelling

In the transition from the air-gap to the AMC-based R-HWP the periodic modulation efficiency changed from sinusoidal to a periodic top-hat shaped function (Fig. 4). This implies an increase of the relative bandwidth of the peaks. For example the fundamental peak has efficiency greater than 90% across a bandwidth of the order of 150%. This means that is possible to actually use just the first peak over such a large bandwidth simply by increasing the fundamental frequency $\nu_0$, i.e. by reducing the layer thicknesses.

The first layer has the lowest refractive index. This can be achieved with a material used for anti-reflection coatings, porous PTFE, which has a refractive index $n_1 \cong 1.25$ at millimetre wavelengths. Going into the practical difficulties in manufacturing a device with the forth layer being an air-gap, we decided to use polypropylene, which has a refractive index of $n_4 \cong 1.50$. This value is still smaller than the one achieved by the last layer and it is enough to guarantee the functioning of the AMC. Using the TL models a new design was optimised to work around the frequency $\nu_0 = 240$ GHz and to provide the refractive indices for the other two quarter-wavelength layers, which resulted to be $n_2 \cong 1.8$ and $n_3 \cong 3.0$.

### B. Artificial dielectric layers design

Two out of the four layers are made with low loss materials which are available at millimetre wavelengths. The second and the third layer require specific refractive indices which are not necessarily available. Using the mesh technology it is possible to engineer artificial dielectrics and a continuum of refractive indices can be achieved by embedding capacitive grids within polypropylene. The distance between the grids must be much smaller than the usual distances used in interference filters, meaning that they cannot be treated as individual lumped elements but rather as a whole structure which increases the refractive index of the supporting medium. This technique has been successfully used in the past to realise anti-reflection coatings [11].

The second and third layers have been designed individually using commercial Finite-element Analysis (FEA) software [9]. The goal was to design two quarter-wavelength thick artificial slabs with refractive indices $n_2 \cong 1.8$ and $n_3 = 3.0$ starting from the lower index polypropylene substrate with $n_{PP} = 1.5$. The model consisted of a unit cell periodic structure with period g much smaller than the minimum operational wavelength (g=100 μm). The mesh-loaded polypropylene slab was sandwiched between two semi-infinite slabs with the generic refractive index that we wanted to mimic (see inset Fig. 6). During the optimization procedure the number of grids, their geometry and their spacing were varied in order to minimize the reflection coefficient Γ, across the whole band. As shown in Fig.5 the final reflection coefficients were both below -28 dB across the 10-490 GHz frequency range, meaning that the two artificial slabs were almost indistinguishable from homogeneous slabs with the targeted refractive indices.

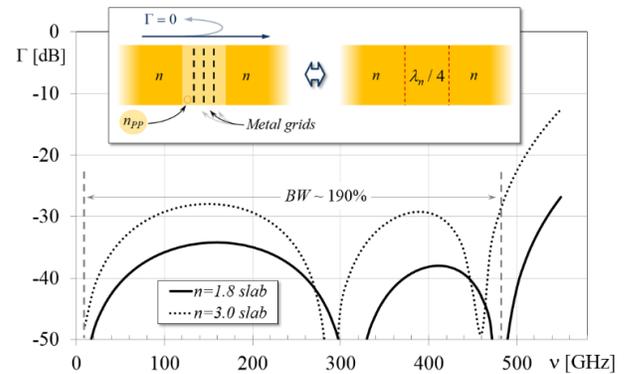

Fig. 5. Artificial dielectric layers model sketch and simulation results.

### C. Embedded R-HWP finite element design

Although the TL modelling provides a very good approximation of the optical performance of systems like the one we are designing, a more detailed and accurate FEA model was required in order to include the details of the metamaterial structures. This FEA simulation only required a unit cell where homogeneous materials were used for the first and fourth layers, whilst the two optimized artificial dielectric layers had all their grid details inserted. The wire-grid was located within the high-to-low index interface. All the grids were modelled as copper with finite-conductivity. The frequency dependence of the polypropylene refractive index and loss tangent were taken into account. Master and slave periodic boundary conditions were used on the sides of the models to simulate infinitely extended slabs, thereby allowing also off-axis incidence studies. The electromagnetic sources were plane-waves that could have arbitrary angles of incidence. We have defined the angle of incidence of the radiation ϑ, the ER-HWP rotation angle α (null when the wire-grid is vertical) and the *S* and *P* polarizations (respectively perpendicular and parallel to the plane of incidence) as sketched in Fig. 6.

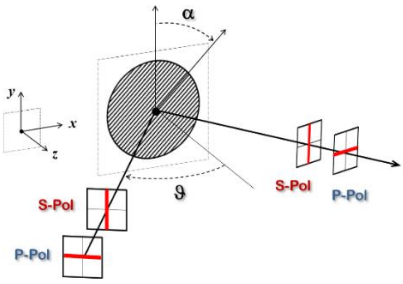

Fig. 6. Polarization and rotation angle definitions used in the ER-HWP models and tests.

The results of the finite-element on-axis simulations ($\vartheta=0°$) are reported in Fig. 7. Although the model includes now also the metamaterial layers the expected performance is still close to ideal. Across the 76 - 383 GHz frequency range, corresponding to a fractional bandwidth of ~ 134%, the modulation efficiency is always above 90% and has an average value of 0.985. The efficiency calculation takes also into account the losses and not just the phase. Within the same spectral region the differential phase-shift is kept within $176° \pm 14°$ and along the two HWP axes ($\alpha=0°, 90°$) the averaged reflection and absorption coefficients are respectively of the order of 0.98 and 0.02

The finite-element model allowed the study of the reflective plate performance when the electromagnetic waves are incident at off-axis angles. In this cases it is not only necessary to distinguish between $S$ and $P$ polarizations but also in respect to the plate rotation angle $\alpha$. We have modelled incidence angles of $\vartheta=22.5°$ and $\vartheta=45°$. The results will be discussed together with the associated measurements in Sec. 5.

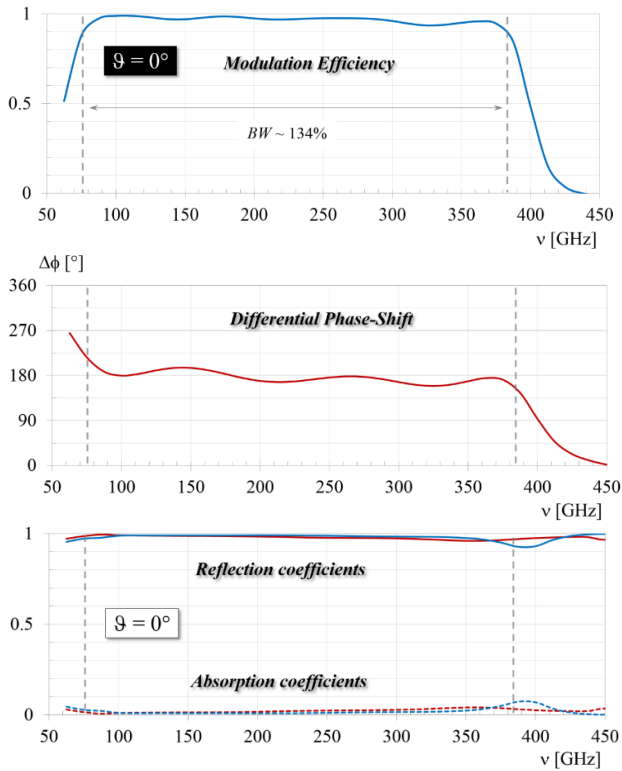

Fig. 7. On-axis ER-HWP finite element simulation results: (a) Reflection and absorption coefficients; (b) Differential phase-shift; (c) Modulation efficiency.

## 4. Manufacture

The device manufacture relies on the long-established metal-mesh production facilities at Cardiff University. Details of the hot-pressed metal-mesh technology have been reported in (Ade et al. SPIE) and rely on lithographic and polymer-processing techniques to embed multiple mesh layers in precisely spaced dielectric substrates. The technology has been used for over 30 years for the production of quasi optical filters and polarizers for astronomical telescopes [12, 13].

The 200 mm diameter prototype device consisted of 4 patterned capacitive copper layers embedded in a 0.5 mm thick polypropylene support. The back surface was a 10 μm period polarizer separated from an inductive reflective layer. The device is shown in Fig. 8; it appears opaque due to the porous PTFE antireflection coating layer applied to the front surface as the final manufacture step.

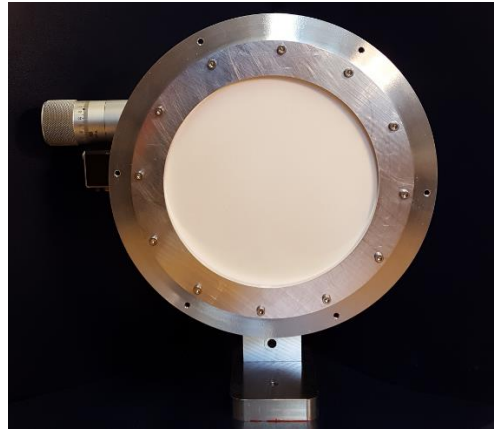

Fig. 8. Picture of the ER-HWP prototype, mounted in a rotary stage, tested in this work.

## 5. Experimental characterization

### A. Experimental configurations

To measure the spectral performance of the ER-HWP we used two Fourier Transform (FT) spectrometers: a Martin-Pupplett polarising FTS [Fig. 9(a)] and a Mach-Zehnder FTS with the addition of two fixed polarisers [Fig. 9(b)].

The Martin-Puplett FTS uses a polarizing beam divider to achieve very large spectral coverage and contains fixed input and output polarisers so the beam incident at the ER-HWP has a fixed polarization state. However, rotation of the output polariser selects the orthogonal port of this spectrometer thus allowing both $S$ and $P$ measurements of the ER-HWP. Standard FT algorithms are used to get the reflection coefficients for each measurement. As a reference we replace the ER-HWP with a copper disc which allows determination of the spectral throughput of the FTS. A ratio of the device spectra against this background provides the inherent device reflectivity.

The Mach-Zehnder spectrometer uses intensity metal mesh beam divider which covers about 4 octaves. For the instrument as configured here we optimised the response for the 120 – 480GHz region. To enable measurement of the ER-HWP we added two fixed polarisers as shown in Fig. 9(b) before and after the test plate. The simultaneous rotation of the two polarizers allowed tests with S and P polarizations.

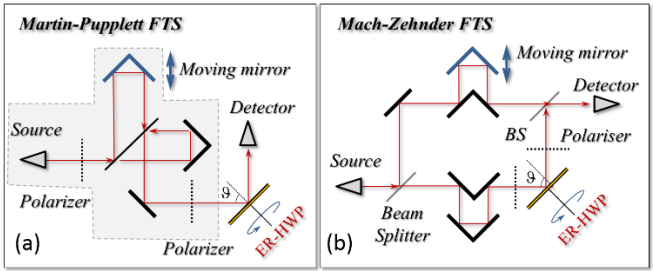

Fig. 9. a) Martin-Puplett polarising FTS, b) Mach-Zehnder FTS with two fixed polarisers. The placement of the ER-HWP is shown for both.

## B. Measurements and results

The ER-HWP was characterised at two incidence angles, $\vartheta=22.5°$ and $\vartheta=45°$, using the Martin-Puplett setup shown in Fig. 9(a). In both cases two parallel wire-grid polarizers, located at the entrance and at the output of the interferometer, were able to define the polarization status of the incident radiation: $S$ when the wires were horizontal and $P$ when vertical. In each of the four previous configurations the waveplate reflection coefficients were acquired for three different rotation angles: $\alpha=0°$, $45°$, $90°$. These allowed characterization of the waveplate axes and its 45 degrees cross-polarization leakage.

The results of the $\vartheta=22.5°$ and $\vartheta=45°$ measurements are reported respectively in Fig. 10 and Fig. 11. In both cases, the reflection coefficients for the $S$ and $P$ polarization (a, b), the polarization modulation efficiency (c) and the differential phase-shift (d) as a function of frequency are superimposed to the finite element models expected performances.

The reflection measurements with the waveplate orientations $\alpha=0, 90°$ provide to the waveplate reflection coefficients along its axes. These are interchanged between the $S$ and $P$ polarizations. The reflection coefficients are above 90% across about two octaves for both incidence angles and polarizations as expected by the relative model predictions. In the $\alpha=45°$ case the effect of the waveplate is to rotate the incoming linear polarization by 90° and, given the alignment of the two polarizers in the FTS, the signal is expected to be at its minimum. Using either the Jones or the Stokes formalism it is possible to show that the reflection coefficients is given by:

$$R_{45} = \frac{1}{4}(R_0 + R_{90} + 2\sqrt{R_0 \cdot R_{90}}\cos\Delta\phi) \quad (1)$$

In the case of identical reflection coefficients along the axes ($R_0 = R_{90}$) and ideal differential phase-shift ($\Delta\phi = \pi$) we would not expect to detect any signal in the original direction. However, the observed cross-polarization leakage gives us information about the polarization modulation efficiency that we can define as follow:

$$\varepsilon = R_0 - R_{45} \quad (2)$$

This equation, valid for any $\vartheta$, $S$ and $P$, is telling us what fraction of the signal, initially aligned and reflected along one axis, is then rotated by 90° when the plate axes are positioned at 45° from the incoming polarization. The modulation efficiencies for the $S$ and $P$ polarizations are different but still very similar. We report the average of them in Fig. 10(c) and Fig.11(c). We see that the modulation efficiencies are above 90% across roughly two octaves and that the bandwidth gets larger by increasing the angle of incidence $\vartheta$.

FTS measurements carried out with a standard setup like the one sketched Fig. 9(a) do not provide phase information directly. In order to extract the differential phase-shift between the waveplate's axes we need to combine the previous three measurements. This can be done by rearranging equation (2) and obtaining:

$$\Delta\phi = \pm\arccos\left[\frac{2R_{45}-0.5(R_0+R_{90})}{\sqrt{R_0\,R_{90}}}\right] \quad (3)$$

The differential phase-shift for the $\vartheta=22.5°$ measurements extracted using equation (3) are reported in Fig. 10(d). Given the *arccos* function sign indeterminacy the phase can appear either all above or all below 180°. Knowing the model predictions we have manually attributed the negative sign to the data below 175 GHz (see dotted line in Fig. 10(d)). We notice that in this method any random noise detected by the FTS at $\alpha=45°$ results in a systematic error which forces the phase to depart from 180°.

For the above reasons in the $\vartheta=45°$ case we have used a completely different experimental setup to measure the differential phase-shift. The modified Mach-Zehnder FTS described earlier and sketched in Fig. 9(b) allows direct measurements of the phase without any sign indeterminacy. In addition, it does not rely on the $\alpha=45°$ cross-polarization measurements with the associated noise problems discussed above. The differential phase-shift is obtained directly by computing the complex Fourier Transforms of the inteferograms at $\alpha=0°$ and $\alpha=90°$, by taking their frequency dependent arguments and by subtracting them, as described in more details in [8]. The differential phase-shifts computed for both the $S$ and $P$ polarisations resulted to be very close to the model predictions across two octaves (Fig. 11(d)).

## 5. Conclusions

We have developed a new type of reflective HWP based on metamaterials. The working principle is completely different from any previous design and it is based on the different phase-shift experienced by two orthogonal polarizations respectively reflecting off an electric conductor and off an artificial magnetic conductor.

A prototype working across the frequency range ~100-400 GHz was manufactured using the mesh-filters technology and tested with two Fourier Transform Spectrometers. The polarisation modulation efficiency results to be greater than 90% over bandwidths of the order of two octaves. The ER-HWP shows good angular stability by maintaining similar performances across the same operational bandwidth up to 45° incidence angles. The differential phase-shift oscillates around 180°, within ±20° and ±30° respectively for the $\vartheta=22.5°$ and $\vartheta=45°$ incidence angles.

The above performance has never been achieved before and exceeds the limits of both the birefringent multi-plate Pancharatnam designs and the transmissive mesh HWPs. Larger bandwidths can be achieved by increasing the number of layers in the gradient index section and also by increasing the value of the refractive index of the last layer.

Future applications of this new device could be found in ground-based, balloon-borne astronomical instrumentation and in the next generation of CMB satellites missions.

**Funding Information.** European Space Agency (ESA) contract number 4000107865/13/ML/MH.

**Acknowledgment**.
The Authors acknowledge support from the European Space Agency through a Technology Research Program.

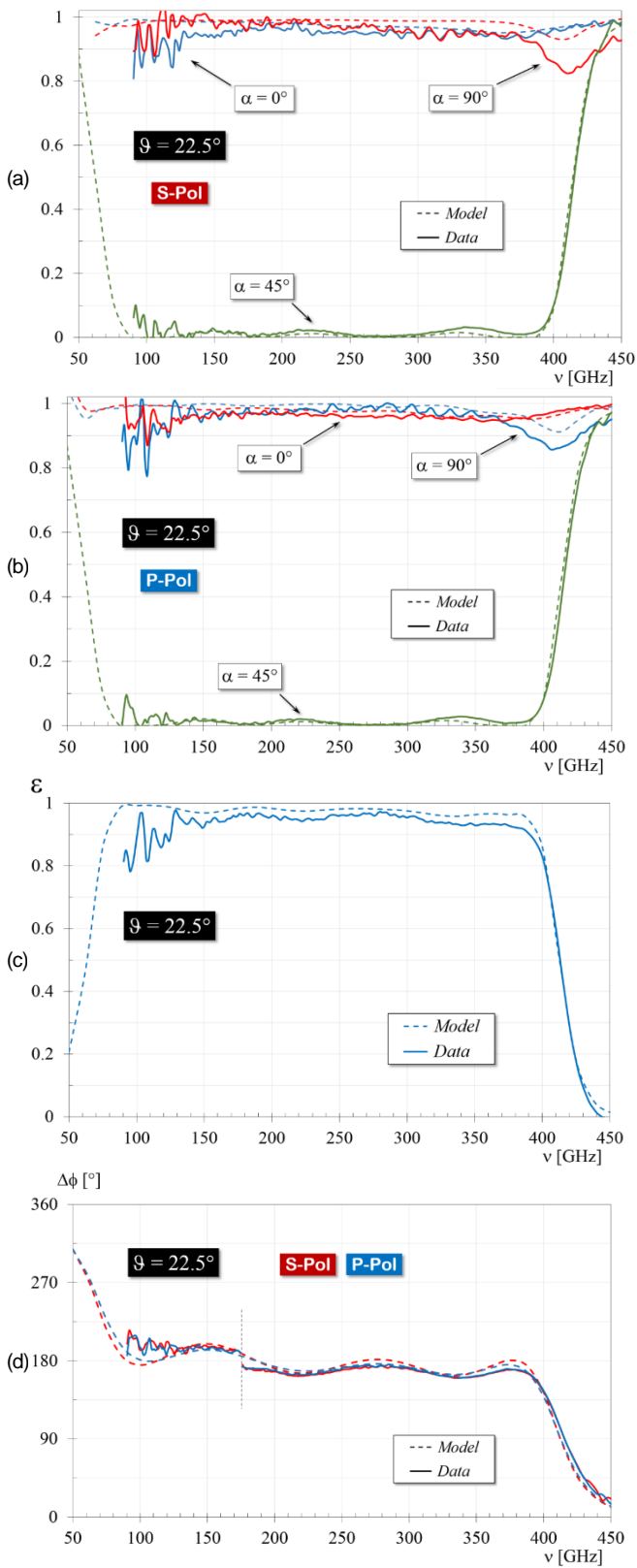

Fig. 10. Off-axis measurements vs model predictions at ϑ=22.5° incidence angle. (a, b) S- and P-polarization reflection coefficients at α=0°, 45°, 90° rotation angles; (c) Averaged S- and P-polarization modulation efficiency. (d): Martin-Pupplett FTS differential phase-shifts.

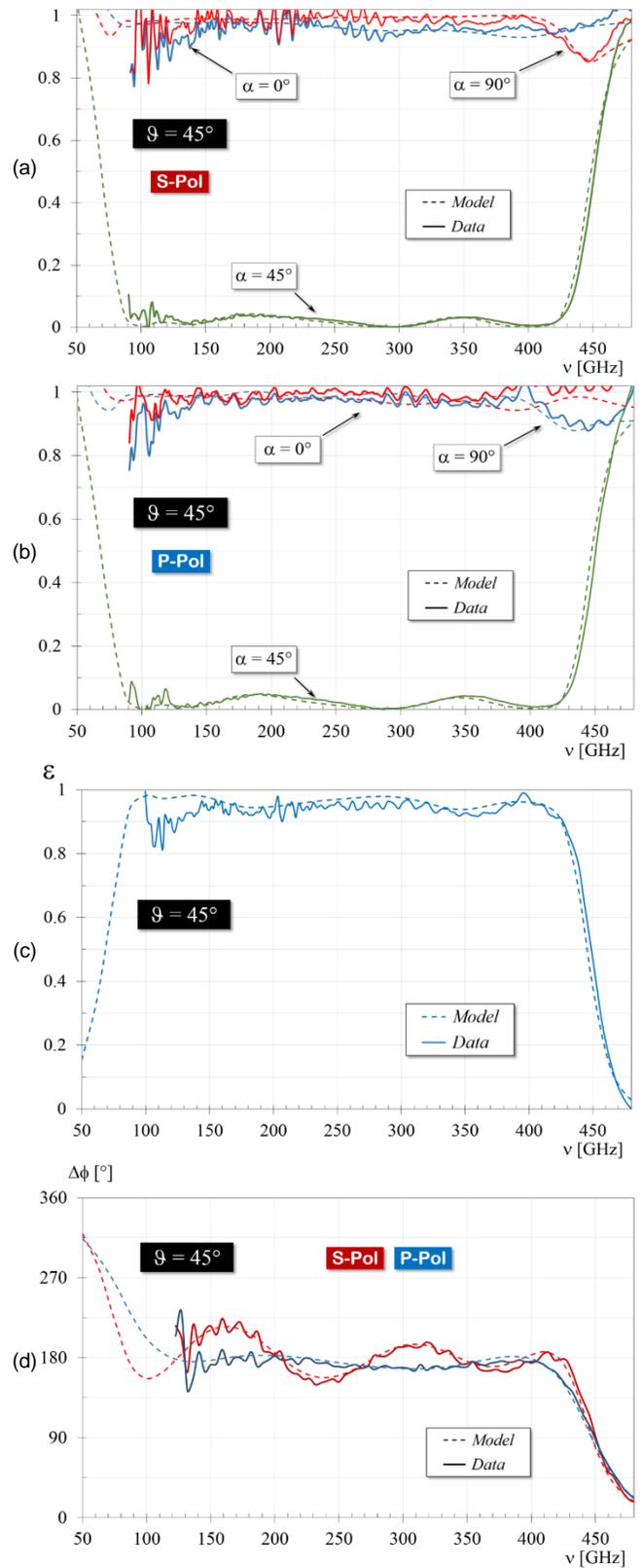

Fig. 11. Off-axis measurements vs model predictions at ϑ=45° incidence angle. (a, b) S- and P-polarization reflection coefficients at α=0°, 45°, 90° rotation angles; (c) Averaged S- and P-polarization modulation efficiency. (d): Mach-Zehnder FTS differential phase-shifts.

## References

1. The COrE Collaboration, "COrE (Cosmic Origins Explorer) A White Paper", arXiv:1102.2181 (2011)
2. G. Pisano, G.Savini, P.A.R.Ade, V.Haynes and W.K. Gear, "Achromatic Half-Wave Plate for Submillimetre Instruments in CMB Astronomy: Experimental Characterisation," Appl. Opt. **45**, 6982-6989 (2006).
3. G. Pisano, G. Savini, P.A.R. Ade, V. Haynes, "A Metal-mesh Achromatic Half-Wave Plate for use at Submillimetre Wavelengths," Appl. Opt. **47**, 6251-6256 (2008).
4. G.Pisano, M. W. Ng, V. Haynes and B. Maffei, "A Broadband Metal-Mesh Half-Wave Plate for Millimetre Wave Linear Polarisation Rotation," Progress In Electromagnetics Research M, **25**, pp.101-114 (2012)
5. G. Pisano, B. Maffei, M.W. Ng, V.Haynes, M. Brown, F. Noviello, P. de Bernardis, S. Masi, F. Piacentini, L. Pagano, M. Salatino, B. Ellison, M. Henry, P. de Maagt, B. Shortt, "Development of Large Radii Half-Wave Plates for CMB satellite missions," Proceedings of the SPIE, **9153**, id. 915317 (2014).
6. P. A. R. Ade, G. Pisano, C. E. Tucker, S. O. Weaver, "A Review of Metal Mesh Filters," Proceedings of the SPIE, **6275**, pp.U2750 (2006).
7. G. Pisano, P. Ade, C. Tucker, P. Moseley and M.W. Ng, "Metal mesh based metamaterials for millimetre wave and THz astronomy applications", Proceedings 8th UCMMT-2015 Workshop Cardiff, pp.1-4 (2016).
8. G. Pisano, P. A. R. Ade, C, Tucker, "Experimental realization of an achromatic magnetic mirror based on metamaterials," Appl. Opt. **55**, 4814-4819 (2016).
9. High Frequency Structure simulator (HFSS): www.ansys.com
10. C.R. Brewitt-Taylor, "Limitation on the bandwidth of artificial perfect magnetic conductor surfaces," IET Micr. Ant. Prop. **1**, 255–260 (2007).
11. Zhang, J., Ade, P.A.R, Mauskopf, P., Moncelsi, L., Savini, G. and Whitehouse, N., "New Artificial Dielectric Metamaterial and its Application as a THz Anti-Reflection Coating," Appl. Opt. **48**, 6635-6642 (2009).
12. P. A. R. Ade, G. Savini,, R. Sudiwala, C. Tucker, A. Catalano, S. Church, R. Colgan, F. X. Desert, E. Gleeson, W.C. Jones, J.-M. Lamarre, A. Lange, Y. Longval, B. Maffei, J. A. Murphy, F. Noviello, F. Pajot, J.-L. Puget, I. Ristorcelli, A. Woodcraft, and V. Yurchenko, "Planck pre-launch status: The optical architecture of the HFI", Astronomy & Astrophysics v.520, A11 (2010).
13. M.J. Griffin et al., "The Herschel-SPIRE instrument and its in-flight performance", Astronomy & Astrophysics, v. 518(4), id.L3 (2010).